\documentclass{article}
\usepackage{spconf,amsmath,graphicx,hyperref}

\usepackage{cuted} 
\usepackage{caption} 

\usepackage{stfloats}

\usepackage{cite}

\def\etal{\emph{et al. }}
\usepackage{amssymb,amsfonts}
\usepackage{algorithmic}
\usepackage{textcomp}
\usepackage{xcolor}
\usepackage{balance}

\usepackage{booktabs}
\usepackage[table]{xcolor}

\usepackage{orcidlink}
\usepackage{svg}
\newcommand{\orcid}[1]{\href{https://orcid.org/#1}{\textcolor[HTML]{A6CE39}{\aiOrcid}}}

\usepackage[linesnumbered,ruled,vlined]{algorithm2e}
\SetKwInput{KwInput}{\hspace{-7pt}Input}                
\SetKwInput{KwModel}{\hspace{-7pt}Model} 
\SetKwInput{KwOutput}{\hspace{-7pt}Output}              
\SetKwInput{KwInitialize}{\hspace{-7pt}Initialize}              
\SetKwInput{KwDefine}{\hspace{-7pt}Define}              
\SetAlFnt{\small}

\SetKwComment{tcp}{$\rhd$\hspace{1pt}}{}

\SetCommentSty{mycommfont}    

\newsavebox\MBox

\def\BibTeX{{\rm B\kern-.05em{\sc i\kern-.025em b}\kern-.08em
    T\kern-.1667em\lower.7ex\hbox{E}\kern-.125emX}}
\begin{document}

\title{Sonic Palette: Predicting Dynamic Color Themes from Music}

%
\name{Wen-Li Wei$^{\star}$ \qquad Kenneth Yang$^{\star \dagger}$ \qquad Jen-Chun Lin$^{\star}$}
\address{$^{\star}$Institute of Information Science, Academia Sinica \\ $^{\dagger}$Graduate Institute of Communication Engineering, National Taiwan University}

%
%
%
%
\maketitle

\begin{abstract}
We present \textit{Sonic Palette}, a music-driven framework for predicting time-varying color themes from audio. Rather than treating music-to-palette generation as a static problem, our method formulates music--color correspondence as a cross-modal sequence prediction task that models how palette-level visual attributes evolve over time in response to music. Specifically, \textit{Sonic Palette} uses a music-conditioned autoregressive Transformer to predict frame-wise color themes, each represented by five dominant colors, from input audio features. To support this task, we curate a dataset of paired audio--color sequences derived from $265$ full-length official music videos. We further visualize the predicted color themes through a palette-guided recoloring process that preserves image structure while revealing temporally varying color appearance associated with the music. Quantitative results show that \textit{Sonic Palette} predicts color themes that more closely match the ground truth, while subjective results indicate more natural temporal color transitions and stronger perceived music--color atmosphere alignment, highlighting its potential for music-driven visual design, computational art, and creative visual authoring.
\end{abstract}
%
\begin{keywords}
Dynamic color theme prediction, Music--color correspondence, Cross-modal sequence prediction
\end{keywords}
\section{Introduction}

Music is inherently temporal, yet its perceptual impact extends beyond the auditory domain. Musical content can evoke vivid visual impressions of color, brightness, and atmosphere, and such cross-modal associations have long influenced artistic practice and perceptual research. For example, gentle passages may suggest muted or warm palettes, whereas energetic passages may evoke more vivid and rapidly changing colors. Prior perceptual studies have provided evidence that music and color are systematically related through affective and acoustic characteristics~\cite{Palmer2013,lindborg2015colour,whiteford2018color,Reymore_2025}. However, recent computational work on music--color modeling has focused on predicting a static palette representation from a music input~\cite{Hu25Music2Palette}. Such a formulation does not capture how color themes evolve as the musical content changes over time.

This limitation is particularly relevant for applications in which music and visual appearance unfold jointly over time. A single palette may characterize the overall color impression of a piece, but cannot capture local variations as the music evolves. Modeling such dynamic music--color correspondence therefore involves not only learning cross-modal associations, but also capturing temporal dependencies across successive color themes. To address this problem, we propose \textit{Sonic Palette}, a music-driven framework for predicting time-varying color themes from audio.

\begin{figure*}
\centering
\includegraphics[width=.99\linewidth]{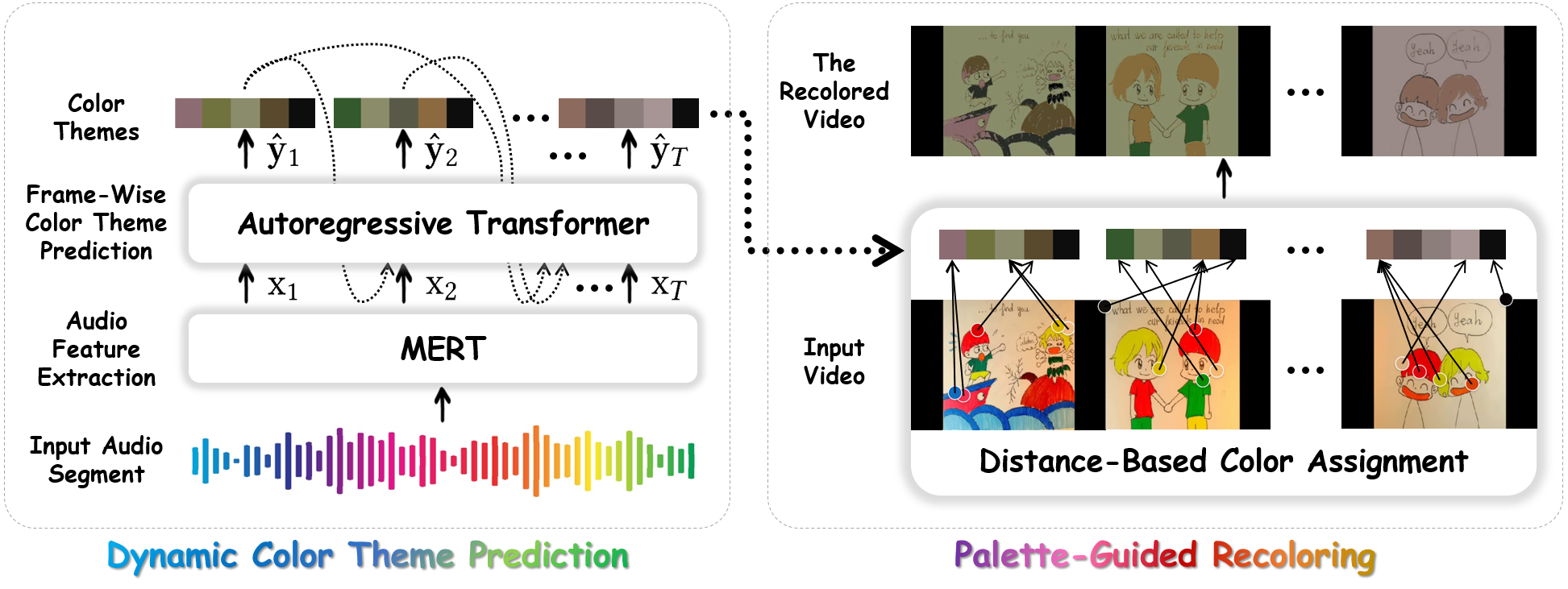}\vspace{-9pt}
  \captionof{figure}{Overview of \textit{Sonic Palette}. Given an input audio segment, the framework predicts a time-varying sequence of color themes to drive a palette-guided recoloring process.}\vspace{-3pt}
  \label{fig:teaser}
\end{figure*}

Rather than targeting detailed or photorealistic image colorization, \textit{Sonic Palette} focuses on palette-level visual attributes that compactly characterize color appearance. We formulate music--color modeling as a cross-modal sequence prediction problem: given a temporally ordered sequence of audio features, the model predicts a corresponding sequence of frame-wise color themes, each represented by five dominant colors. The predicted sequence can subsequently be used in a palette-guided recoloring process to visualize how music-conditioned color appearance evolves over time (see Figure~\ref{fig:teaser}). Unlike prior music-to-palette generation work that produces a single static palette representation for a music input~\cite{Hu25Music2Palette}, our formulation explicitly models the temporal evolution of color themes throughout the music.

To support this task, we curate a dataset of $265$ full-length official music videos and derive temporally aligned audio--color sequences from them. Each video frame is represented by a color theme consisting of five dominant colors~\cite{kobayashi1991color}, providing a structured representation of its palette-level appearance. Audio representations are extracted using the pre-trained MERT model~\cite{li2024mert}. We then train a music-conditioned model based on an autoregressive Transformer~\cite{chen2020generative,zhang2023t2mgpt,SA2025CamSketch,Zhu2026TaiChi} to predict the frame-wise color-theme sequence. By conditioning each prediction on both the temporally contextualized audio features and preceding color themes, the model captures dependencies across successive color themes and their evolution in response to music. This autoregressive formulation encourages coherent transitions rather than treating each frame-wise prediction independently.

Our contributions are threefold. First, we introduce a dataset of paired audio--color sequences derived from $265$ full-length official music videos, providing a resource for studying music-driven dynamic color theme prediction. Second, we formulate music-to-palette generation as a cross-modal sequence prediction problem, extending static palette generation toward temporally evolving frame-wise color themes. Third, we quantitatively evaluate the predicted color-theme sequences and conduct subjective evaluations of their palette-guided recoloring results, showing closer agreement with the ground truth, more natural temporal color transitions, and stronger perceived music--color atmosphere alignment. Our results highlight the potential of dynamic music--color modeling for music-driven visual design, computational art, and creative visual authoring.

\vspace{-3pt}
\section{Related Work}
\subsection{Music--Color Correspondence}

Palmer \etal\cite{Palmer2013} found systematic music--color associations that were strongly mediated by perceived emotion; for example, faster major-mode music tends to be associated with lighter, more saturated, and yellower colors, whereas slower minor-mode music tends to be associated with darker, less saturated, and bluer colors. Lindborg and Friberg~\cite{lindborg2015colour} further examined this relationship using an interface for navigating the CIE Lab color space and showed that models incorporating perceived emotion explained 60--75\% of the variation in color responses, providing further evidence for emotion-mediated music--color associations. Whiteford \etal\cite{whiteford2018color} subsequently found systematic music--color associations across a broader range of musical styles, further supporting the role of emotional mediation. More recently, Reymore and Lindsey~\cite{Reymore_2025} examined musical timbre--color correspondences, showing that semantic timbre ratings predicted the lightness and saturation of matched colors. Collectively, these studies provide a perceptual foundation for computational modeling of music--color correspondence.\vspace{-7pt}

\subsection{Music-to-Palette Generation}

Building on perceptual evidence of music--color correspondence, recent computational work has explored automatic color generation from music. Music2Palette~\cite{Hu25Music2Palette} formulates music-to-palette generation as a cross-modal representation learning problem and constructs emotion-aligned music--palette pairs for direct palette generation from music. Its framework jointly promotes emotion alignment, color diversity, and palette coherence, demonstrating the feasibility of learning direct mappings between music and color palettes. However, Music2Palette produces a single static palette representation for a music input rather than explicitly modeling the temporal evolution of color themes throughout a full-length piece. In contrast, our work formulates music--color correspondence as a cross-modal sequence prediction problem, predicting frame-wise color themes that evolve over time in response to music.\vspace{-8pt}

\section{Approach}\vspace{-3pt}

Given paired training data $(\mathbf{X}, \mathbf{Y})$, where $\mathbf{X}=\{\mathbf{x}_t\}_{t=1}^{T}$ denotes a sequence of MERT audio features and $\mathbf{Y}=\{\mathbf{y}_t\}_{t=1}^{T}$ denotes the corresponding ground-truth frame-wise color themes, our goal is to model the temporal correspondence between musical content and palette-level visual appearance. We train a music-conditioned autoregressive Transformer to predict a sequence of frame-wise color themes from audio features over time, as illustrated in Figure~\ref{fig:teaser}. Each audio feature $\mathbf{x}_t \in \mathbb{R}^{F}$ encodes the musical content associated with frame $t$, and the input sequence is organized as $\mathbf{X}\in\mathbb{R}^{T\times F}$, where $T$ and $F$ denote the sequence length and feature dimension, respectively. By retaining the temporal ordering of both modalities, the formulation enables the model to learn how successive color themes vary with the corresponding audio context. During training, the audio and its corresponding color-theme sequence are divided into non-overlapping $2$-second segments, with each pair $(\mathbf{X}, \mathbf{Y})$ corresponding to one such segment. At inference time, the input audio is processed in the same segment-wise manner.

\begin{figure}[t]%
    \centering{\includegraphics[width=\linewidth]{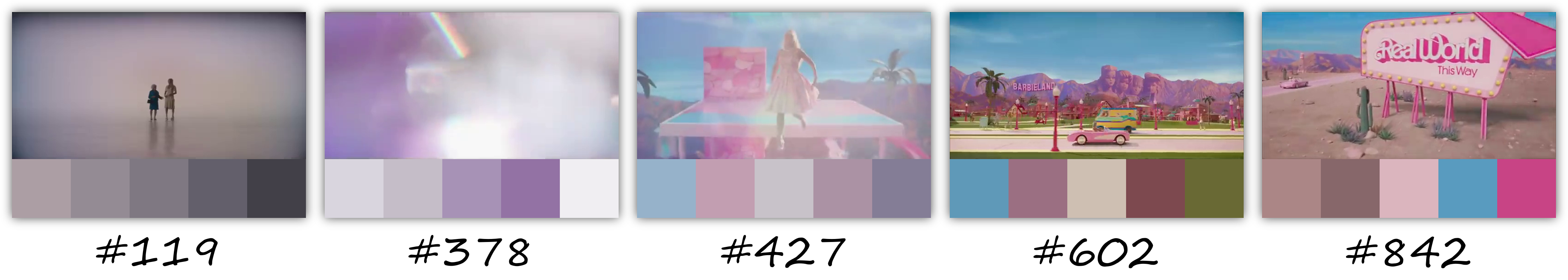}}\vspace{-7pt}
    \caption{Examples of frame-wise color themes from Billie Eilish's ``What Was I Made For?'' (from \textit{Barbie}).}
    \label{fig:color_theme}%
\end{figure}

To obtain a compact and structured output space, we represent each frame-wise color theme $\mathbf{y}_t$ using five dominant colors~\cite{kobayashi1991color,lin2016automatic}. The dominant colors are selected from a quantized RGB histogram in descending order of color frequency, while Euclidean distance-based filtering is applied to avoid selecting near-duplicate colors (see Figure~\ref{fig:color_theme}). This representation summarizes the major color composition of a frame while avoiding the complexity of directly predicting dense pixel-level appearance. Each dominant color is represented by three RGB channels, with each channel quantized into $16$ discrete bins. The quantization therefore converts color prediction from continuous RGB regression into a set of discrete classification problems, yielding a compact palette-level output space for dynamic color theme prediction. Concretely, each frame-wise color theme $\mathbf{y}_t$ is represented as a $5\times3$ matrix of discrete bin labels, or equivalently as a $5\times3\times16$ one-hot tensor.

Under this formulation, the model predicts the conditional distribution of $\mathbf{y}_t$ as $p(\mathbf{y}_t \mid \mathbf{y}_{1:t-1}, E(\mathbf{X}))$, where $\mathbf{y}_{1:t-1}$ denotes the preceding color themes and $E(\mathbf{X})$ denotes the sequence of temporally contextualized audio features obtained by applying a Transformer encoder~\cite{vaswani2017attention} to the input sequence of MERT audio features. The training objective aggregates cross-entropy losses over the discrete bin labels of all RGB channels across all dominant colors. At inference time, the highest-probability bin is selected for each RGB channel of each dominant color to form the predicted color theme $\hat{\mathbf{y}}_t$. Previously predicted color themes are autoregressively fed back to predict subsequent themes (see Figure~\ref{fig:teaser}). Repeating this autoregressive process over time yields a predicted sequence of frame-wise color themes of length $T$.

To provide a perceptual visualization of the predicted dynamic color themes, we further apply a palette-guided recoloring process, as shown in Figure~\ref{fig:teaser}. Given an input image and a predicted color theme $\hat{\mathbf{y}}_t$, let $\hat{\mathbf{c}}_{t,k}\in\mathbb{R}^{3}$ denote the quantized RGB color obtained by mapping the predicted bin labels $\hat{\mathbf{y}}_{t,k}$ of the $k$-th dominant color back to RGB values. Each pixel $\mathbf{p}$ in the input image is then assigned to its nearest predicted dominant color according to Euclidean distance in RGB space:\vspace{-5pt}
\begin{equation}
k^{*}
=
\arg\min_{k\in\{1,\ldots,5\}}
\left\|
\mathbf{p}-\hat{\mathbf{c}}_{t,k}
\right\|_{2}.\vspace{-5pt}
\end{equation}
The pixel is subsequently recolored using $\hat{\mathbf{c}}_{t,k^{*}}$. Applying this mapping to all pixels preserves the spatial layout and underlying image structure while replacing the original color appearance according to the predicted color theme. Repeating this recoloring process over successive predicted themes provides a direct visualization of how the music-conditioned color appearance changes over time.

\section{Experiments}
\subsection{Dataset and Evaluation Metrics}
We curate a dataset of $265$ full-length official music videos, comprising approximately $18.5$ hours of audiovisual content and $1.7$M video frames, and derive temporally aligned audio--color sequences for studying music-driven dynamic color theme prediction. Among these videos, $65$ are drawn from the DEAP dataset~\cite{Koelstra12DEAP}, while the remaining $200$ are collected from online sources to broaden the musical and visual coverage of the dataset. To evaluate the model under diverse emotional conditions, we construct the validation and test sets from the DEAP subset, which provides valence--arousal annotations for each video. In all experiments, we use $225$ videos for training, $20$ for validation, and $20$ for testing. Both validation and test splits are balanced across four valence--arousal quadrants, with five videos selected from each quadrant, enabling evaluation across a range of affective conditions. Upon acceptance, we will release source links for the official music videos, along with frame-wise color-theme annotations, to support reproducibility and future research.

We evaluate color-theme prediction quality using \emph{Average Color Bin Difference} (ACBD). Since the five dominant colors in each color theme are ordered by color frequency, we compare the predicted and ground-truth dominant colors at the same rank. Let $\hat{\mathbf{y}}_{t,k}=(\hat{y}^{R}_{t,k},\hat{y}^{G}_{t,k},\hat{y}^{B}_{t,k})$ and $\mathbf{y}_{t,k}=(y^{R}_{t,k},y^{G}_{t,k},y^{B}_{t,k})$ denote the predicted and ground-truth RGB bin labels of the $k$-th dominant color at frame $t$, respectively. We define the color bin difference for a single dominant color as
\begin{equation}
d_{t,k}=\frac{1}{3}\sum_{c\in\{R,G,B\}} \left|\hat{y}^{c}_{t,k}-y^{c}_{t,k}\right|.
\end{equation}
We then average over all five dominant colors and all frames within a segment:
\begin{equation}
\mathrm{ACBD}=\frac{1}{5T}\sum_{t=1}^{T}\sum_{k=1}^{5} d_{t,k}.
\end{equation}
Lower ACBD indicates smaller discrepancies between the predicted and ground-truth color themes. We report the average ACBD over all test segments.

We also measure temporal variation using \emph{Average Temporal Bin Difference} (ATBD):
\begin{equation}
\delta_{t,k}
=
\frac{1}{3}\sum_{c\in\{R,G,B\}}
\left||
\hat{y}^{c}_{t,k}-\hat{y}^{c}_{t-1,k}|
-
|y^{c}_{t,k}-y^{c}_{t-1,k}|
\right|,
\end{equation}
\begin{equation}
\mathrm{ATBD}
=
\frac{1}{5(T-1)}
\sum_{t=2}^{T}\sum_{k=1}^{5}\delta_{t,k}.
\end{equation}
Lower ATBD indicates closer agreement between the predicted and ground-truth temporal variations. We report the average ATBD over all test segments.

\begin{table*}[t]
\centering
\small
\renewcommand{\arraystretch}{1.0} 
\setlength{\tabcolsep}{6pt}    
\caption{Quantitative Evaluation of Music-Driven Dynamic Color Theme Prediction}\vspace{-5pt}
\label{tab:evaluation_results}
\begin{tabular}{lcc}
\toprule
\rowcolor{gray!10} \textbf{Methods} & \textbf{Average ACBD} $\downarrow$ & \textbf{Average ATBD} $\downarrow$ \\
\midrule
Static MLP & 4.6757 & 0.9959 \\
Non-autoregressive Transformer & 3.8853 & 1.0068 \\
\rowcolor{blue!5} 
\textbf{Ours} & \textbf{3.5161} & \textbf{0.8671} \\
\bottomrule
\end{tabular}\vspace{10pt}
\end{table*}

\begin{figure*}[t]%
    \centering{\includegraphics[width=.95\linewidth]{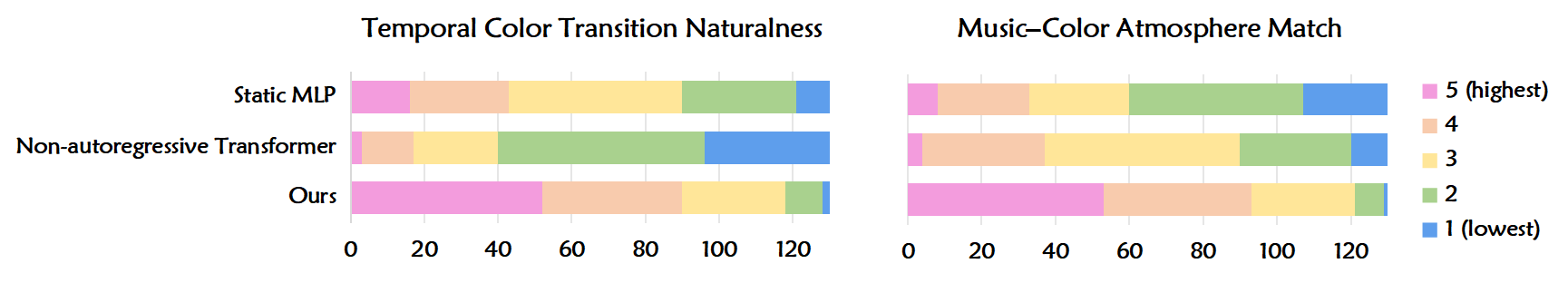}}\vspace{-9pt}
    \caption{Subjective $5$-point opinion scores from $26$ raters.}\vspace{-2pt}
    \label{fig:example}%
\end{figure*}

\subsection{Quantitative and Subjective Results}\vspace{-3pt}

Since existing music-to-palette work is not directly applicable to frame-wise dynamic color-theme prediction, we quantitatively compare our autoregressive Transformer against two task-specific baselines: a non-autoregressive Transformer that predicts each frame-wise color theme using temporally contextualized audio features without conditioning on previously predicted outputs, and a static Multilayer Perceptron (MLP) that predicts a single color theme for an entire song without modeling temporal evolution. All models are optimized using Adam~\cite{kingma2015adam} with dropout~\cite{srivastava2014dropout} applied to mitigate overfitting. The static MLP is trained for $100$ epochs with a learning rate of $5\times10^{-4}$. The non-autoregressive Transformer and our autoregressive Transformer are both trained for $50$ epochs, with learning rates of $1\times10^{-4}$ and $5\times10^{-5}$, respectively. As shown in Table~\ref{tab:evaluation_results}, our method outperforms the non-autoregressive Transformer baseline, achieving lower ACBD and ATBD. These results suggest that conditioning each prediction on preceding color themes improves frame-wise prediction accuracy and better captures temporal variation relative to the ground truth. We also observe clear improvements over the static MLP baseline, consistent with the limitation of predicting a single color theme for an entire song, which cannot represent frame-wise temporal variation. For fair ACBD and ATBD computation, the color theme predicted by the static MLP is repeated across all frames of the song for evaluation.

Subjective evaluation using $5$-point opinion scores across five evaluation sets, each consisting of three recolored videos paired with the same music, further indicates that our method receives higher ratings than both baselines for Temporal Color Transition Naturalness and Music--Color Atmosphere Match (see Figure~\ref{fig:example}). This suggests that our recolored results exhibit more natural temporal color transitions with less perceived flicker and stronger perceived music--color atmosphere alignment. In comparison, the non-autoregressive Transformer tends to produce abrupt frame-to-frame changes in predicted color themes, which can lead to noticeable temporal flicker in the recolored video. The static MLP, by contrast, uses a temporally constant predicted color theme throughout the song and therefore does not explicitly model temporal color-theme variation. Although its recolored appearance may still vary across frames because the recoloring process depends jointly on the input image content and the predicted color theme, such variation arises from changes in the input imagery under a fixed predicted color theme, rather than from temporal adaptation of the color theme itself. Consequently, the static MLP cannot directly adapt its color-theme predictions to temporal changes in the music, limiting its ability to model time-varying music--color atmosphere alignment. These differences are also illustrated in the accompanying demo video\footnote{\url{https://youtu.be/VnEoSkbveXE}}. Overall, the quantitative and subjective results consistently support the effectiveness of our autoregressive formulation for dynamic music--color modeling.\vspace{-8pt}

\section{CONCLUSION}\vspace{-3pt}
We presented \textit{Sonic Palette}, a music-driven framework that formulates music--color correspondence as a cross-modal sequence prediction problem for predicting time-varying color themes from audio. Using a music-conditioned autoregressive Transformer, our method models the temporal evolution of successive frame-wise color themes in response to music. We curate a dataset of $265$ full-length official music videos to support this task. Experimental results show that \textit{Sonic Palette} achieves more accurate color-theme predictions and more closely matches the ground-truth temporal variation, while subjective evaluation indicates more natural temporal color transitions and stronger perceived music--color atmosphere alignment. These results highlight the potential of dynamic music--color modeling for music-driven visual design, computational art, and creative visual authoring.

\vfill\pagebreak

\section{Acknowledgment}
This work was supported by NSTC Grant 113-2221-E-001-017-MY3.

\section{Compliance with Ethical Standards}
This study involved an anonymous and voluntary subjective evaluation of recolored videos, and no personally identifiable information was collected.

\bibliographystyle{IEEEbib}
\bibliography{strings,refs}    

@inproceedings{Hu25Music2Palette,
author = {Jiayun Hu and Yueyi He and Tianyi Liang and Changbo Wang and Chenhui Li},
title = {Music2{P}alette: Emotion-aligned Color Palette Generation via Cross-Modal Representation Learning},
year = {2025},
booktitle = {Proceedings of the 33rd ACM International Conference on Multimedia (MM)}
}

@book{kobayashi1991color,
  title={Color Image Scale},
  author={Kobayashi, Shigenobu},
  year={1991},
  publisher={Kodansha International},
  isbn={9784770015648}
}

@inproceedings{li2024mert,
title={{MERT}: Acoustic Music Understanding Model with Large-Scale Self-supervised Training},
author={Yizhi Li and Ruibin Yuan and Ge Zhang and Yinghao Ma and Xingran Chen and Hanzhi Yin and Chenghao Xiao and Chenghua Lin and Anton Ragni and Emmanouil Benetos and Norbert Gyenge and Roger Dannenberg and Ruibo Liu and Wenhu Chen and Gus Xia and Yemin Shi and Wenhao Huang and Zili Wang and Yike Guo and Jie Fu},
booktitle={International Conference on Learning Representations (ICLR)},
year={2024}
}

@article{Koelstra12DEAP,
author = {Koelstra, Sander and M{\"u}hl, Christian and Soleymani, Mohammad and Lee, Jong-Seok and Yazdani, Ashkan and Ebrahimi, Touradj and Pun, Thierry and Nijholt, Anton and Patras, Ioannis},
title = {{DEAP}: A Database for Emotion Analysis Using Physiological Signals},
year = {2012},
volume = {3},
number = {1},
journal = {IEEE Transactions on Affective Computing},
pages = {18–31}
}

@article{Palmer2013,
  author = {Palmer, Stephen E. and Schloss, Karen B. and Xu, Zoe and Prado-Le{\'o}n, Lilia R.},
  title = {Music--color associations are mediated by emotion},
  journal = {Proceedings of the National Academy of Sciences},
  volume = {110},
  number = {22},
  pages = {8836--8841},
  year = {2013},
  publisher = {National Academy of Sciences},
  doi = {10.1073/pnas.1212562110},
  url = {https://www.pnas.org/doi/10.1073/pnas.1212562110}
}

@article{lindborg2015colour,
  title={Colour Association with Music Is Mediated by Emotion: Evidence from an Experiment Using a {CIE} {L}ab Interface and Interviews},
  author={Lindborg, PerMagnus and Friberg, Anders K.},
  journal={PLoS ONE},
  volume={10},
  number={12},
  year={2015},
  publisher={Public Library of Science},
  doi={10.1371/journal.pone.0144013}
}

@article{whiteford2018color,
  author = {Whiteford, Kelly L. and Schloss, Karen B. and Helwig, Nathaniel E. and Palmer, Stephen E.},
  title = {Color, Music, and Emotion: Bach to the Blues},
  journal = {i-Perception},
  volume = {9},
  number = {6},
  year = {2018},
  doi = {10.1177/2041669518808535},
  URL = {https://doi.org}
}

@article{Reymore_2025,
  author = {Reymore, Lindsey and Lindsey, Delwin T.},
  title = {Color and tone color: audiovisual crossmodal correspondences with musical instrument timbre},
  journal = {Frontiers in Psychology},
  volume = {15},
  number = {1520131},
  year = {2025},
  doi = {10.3389/fpsyg.2024.1520131},
  url = {https://www.frontiersin.org/journals/psychology/articles/10.3389/fpsyg.2024.1520131/full}
}

@inproceedings{chen2020generative,
  title={Generative pretraining from pixels},
  author={Chen, Mark and Radford, Alec and Child, Rewon and Wu, Jeffrey and Jun, Heewoo and Luan, David and Sutskever, Ilya},
  booktitle={Proceedings of the 37th International Conference on Machine Learning (ICML)},
  year={2020}
}

@inproceedings{zhang2023t2mgpt,
  title={{T2M-GPT}: Generating Human Motion from Textual Descriptions with Discrete Representations},
  author={Zhang, Jianrong and Zhang, Yangsong and Cun, Xiaodong and Huang, Shaoli and Zhang, Yong and Zhao, Hongwei and Lu, Hongtao and Shen, Xi},
  booktitle={Proceedings of the IEEE/CVF Conference on Computer Vision and Pattern Recognition (CVPR)},
  year={2023}
}

@inproceedings{SA2025CamSketch,
author = {Wei, Wen-Li and Hu, Yu-Hua and Lin, Jen-Chun},
title = {Cam{S}ketch: Authoring Camera Trajectories for 3{D} Virtual Concerts via Example-Derived Codes and User Sketching},
year = {2025},
booktitle = {Proceedings of the SIGGRAPH Asia 2025 Posters}
}

@article{Zhu2026TaiChi,
  title = {Tai-{C}hi: Text-to-motion generation with locality-aware bipartite body-part motion prior},
  author = {Jian-Kai Zhu and Wen-Li Wei and Jen-Chun Lin},
  journal = {Computers \& Graphics},
  volume = {136},
  year = {2026}
}

@inproceedings{vaswani2017attention,
  title={Attention is all you need},
  author={Vaswani, Ashish and Shazeer, Noam and Parmar, Niki and Uszkoreit, Jakob and Jones, Llion and Gomez, Aidan N. and Kaiser, Łukasz and Polosukhin, Illia},
  booktitle={Advances in Neural Information Processing Systems (NIPS)},
  year={2017}
}

@inproceedings{kingma2015adam,
  author    = {Diederik P. Kingma and Jimmy Ba},
  title     = {Adam: A Method for Stochastic Optimization},
  booktitle = {Proceedings of the 3rd International Conference on Learning Representations (ICLR)},
  year      = {2015}
}

@article{srivastava2014dropout,
  author  = {Nitish Srivastava and Geoffrey Hinton and Alex Krizhevsky and Ilya Sutskever and Ruslan Salakhutdinov},
  title   = {Dropout: A Simple Way to Prevent Neural Networks from Overfitting},
  journal = {Journal of Machine Learning Research},
  year    = {2014},
  volume  = {15},
  number  = {56},
  pages   = {1929--1958},
  url     = {http://jmlr.org}
}

@inproceedings{lin2016automatic,
  author    = {Lin, Jen-Chun and Wei, Wen-Li and Wang, Hsin-Min},
  title     = {Automatic Music Video Generation Based on Emotion-Oriented Pseudo Song Prediction and Matching},
  booktitle = {Proceedings of the 24th ACM International Conference on Multimedia (MM)},
  year      = {2016}
}

\end{document}